\documentclass[%
 reprint,
 amsmath,amssymb,
 aps, superscriptaddress
]{revtex4-1}

\usepackage{eurosym}
\usepackage{amsmath}
\usepackage{graphicx}
\usepackage{dcolumn}
\usepackage{bm}
\usepackage{mathtools}

\begin{document}

\preprint{APS/123-QED}

\title{Long distance magnon transport in the van der Waals antiferromagnet CrPS\textsubscript{4}}
\author{Dennis K. de Wal}\email{d.k.de.wal@rug.nl}
\author{Arnaud Iwens}
\author{Tian Liu}
\address{Zernike Institude for Advanced Materials, University of Groningen,
Groningen, the Netherlands}
\author{Ping Tang}
\address{Advanced Institute for Materials
Research (AIMR), Tohoku University, Sendai, Japan}
\author{ Gerrit E. W. Bauer}
\address{Zernike Institude for Advanced Materials, University of Groningen,
Groningen, the Netherlands} 
\address{Advanced Institute for Materials
Research (AIMR), Tohoku University, Sendai, Japan} 
\address{Kavli Institute
for Theoretical Sciences, University of the Chinese Academy of Sciences,
Beijing, China}
\author {Bart J. van Wees}\address{Zernike Institude for Advanced Materials, University of Groningen,
Groningen, the Netherlands}
\date{\today }

\begin{abstract}
We demonstrate the potential of van der Waals magnets for spintronic
applications by reporting long-distance magnon spin transport in the electrically
insulating antiferromagnet chromium thiophosphate (CrPS\textsubscript{4}) with
perpendicular magnetic anisotropy. We inject and detect magnon spins non-locally by
Pt contacts and monitor the non-local resistance as a function of an in-plane
magnetic field up to 7 Tesla. We observe a non-local resistance over distances
up to at least a micron below the Neel temperature (T\textsubscript{N} = 38
Kelvin)  close to magnetic field strengths that saturate the sublattice
magnetizations.

\end{abstract}
\maketitle

\preprint{APS/123-QED}



\address{Zernike Institude for Advanced Materials, University of Groningen,
Groningen, the Netherlands}

\address{Advanced Institute for Materials Research (AIMR), Tohoku
University, Sendai, Japan}

\address{Zernike Institude for Advanced Materials, University of Groningen,
Groningen, the Netherlands} \address{Advanced Institute for Materials
Research (AIMR), Tohoku University, Sendai, Japan} \address{Kavli Institute
for Theoretical Sciences, University of the Chinese Academy of Sciences,
Beijing, China}

\address{Zernike Institude for Advanced Materials, University of Groningen,
Groningen, the Netherlands}

Since the discovery of the long-range magnetic order in mono- and bilayers of
Cr\textsubscript{2}Ge\textsubscript{2}Te\textsubscript{6}
\cite{gong_discovery_2017} and CrI\textsubscript{3}
\cite{huang_layer-dependent_2017} many (anti)ferromagnetic van der Waals
materials have been identified in monolayer or few layer thicknesses. They are attractive platforms for spintronics due to the rich spin textures caused by the
interplay of inter- and intralayer exchange and magnetic anisotropies.

Many antiferromagnetic van der Waals materials are electrically insulating at
low temperatures, which implies the absence of magnetization damping by free
carriers. They are therefore attractive for the study of collective
excitations of the magnetic order, i.e. spin waves and its quanta, the magnons
\cite{bloch_zur_1930,kittel_introduction_2005}. Magnon transport has been
extensively studied in conventional magnets by, e.g., spin pumping
\cite{tserkovnyak_enhanced_2002}, the spin Seebeck effect (SSE)
\cite{uchida_spin_2010}, and electrical magnon spin injection/detection
\cite{cornelissen_long-distance_2015}. Long distance magnon transport in the
antiferromagnets hematite \cite{lebrun_anisotropies_2019}, nickel
oxide \cite{hahn_comparative_2013}, and YFeO\textsubscript{3}
\cite{das_anisotropic_2022} has been demonstrated. Ultrathin films of the low-damping ferrimagnetic yttrium iron garnet (YIG), the
material of choice for efficient magnon transport,
show the beneficial effects of two-dimensional (2D) \textit{vs}.
three-dimensional (3D) transport in the form of strongly enhanced magnon
conductivities \cite{wei_giant_2022}. Magnon spin transport driven by temperature gradients
(SSE) \cite{rezende_theory_2016} has been reported in ferro- and
antiferromagnetic van der Waals materials
\cite{liu_spin_2020,xing_magnon_2019}. However, the local and non-local SSEs
provide only convoluted information on the magnon transport properties. Thermal magnon currents are generated by thermal gradients in the entire sample, making it difficult to disentangle the magnon relaxation length and magnon spin conductivity \cite{cornelissen_long-distance_2015, wei_giant_2022}.
Antiferromagnetic resonance of CrCl\textsubscript{3}
\cite{macneill_gigahertz_2019} reveals the existence of acoustic and optical
magnon modes, but does not resolve their roles in spin transport.
In order to assess the potential of van der Waals magnets for spintronic
applications, we therefore have to study the propagation of magnons that are locally
generated by microwaves or, as we will show here, by electrical injection.

Heavy metal contacts such as Pt with a large (inverse) spin Hall effect have
become a standard instrument to study magnetic materials. The spin Hall
magnetoresistance (SMR) in a Pt contact is a reliable method to measure the
surface equilibrium magnetization \cite{nakayama_spin_2013}, which has already
been used to study CrPS$_{4}$ \cite{wu_magnetotransport_2022} and FePS\textsubscript{3} \cite{feringa_observation_2022}. With two Pt contacts, the spins injected by an
electric current in one terminal by the spin Hall effect propagate in an
electrically insulating magnet in the form of magnons, which can be detected
by another contact via the inverse spin Hall effect
\cite{cornelissen_long-distance_2015}. Here we report, to the best of our
knowledge for the first time, such non-local electrical measurements of magnon
transport in a van der Waals antiferromagnet, in our case CrPS\textsubscript{4}

\begin{figure*}[ptb]
\centering
\includegraphics[width=0.85\textwidth]{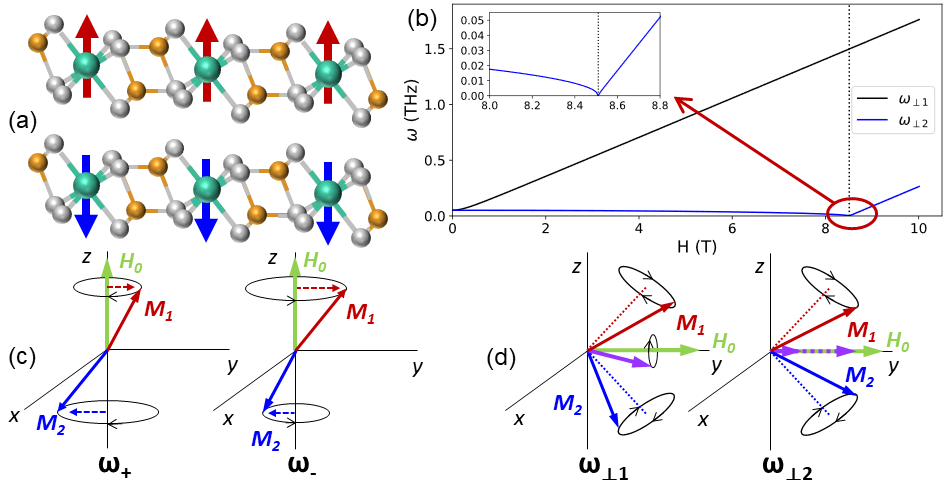}\caption{Spin
texture and magnon modes in antiferromagnetic CrPS \textsubscript{4}. (a) Atom
and spins of a bilayer of CrPS\textsubscript{4}. Red and blue arrows indicate
the local magnetic moments of the Cr atoms (turquoise). The interlayer
(intralayer) exchange coupling is ferromagnetic (antiferromagnetic). (b)
In-plane magnetic field dependence of the magnon band edges. (c) Optical
$(\omega_{+})$ and acoustic magnon modes at out-of-plane (oop) magnetic fields below the
spin-flop transition (d)  magnon modes at in-plane (ip) magnetic fields below the
spin-flip transition $(H<H_{E\perp}).$  The net magnetization of the
$\omega_{\perp1}$ mode precesses (purple vector) around the ip external field
vector with equal modulus, while that of the $\omega_{\perp2}$ mode
oscillates in the direction of the field. }%
\label{fig1:Figure1}%
\end{figure*}

CrPS\textsubscript{4} is an A-type antiferromagnet (see Figure
\ref{fig1:Figure1}(a)). Individual layers are out-of-plane (oop) 2D
ferromagnets, but consecutive layers order antiferromagnetically at a N\'{e}el
temperature $T_{N}\simeq38$ K. Its relative stability in air facilitates the
fabrication of devices. An oop field of $H_{\mathrm{spinflop}}\approx$ 0.9 T (at 5 K) induces a spin-flop
transition to a canted state, while the magnetization becomes saturated into a
\textquotedblleft spin flip" state at 8.5 T. In-plane (ip) fields result in magnetization
saturation at nearly the same field, indicating that the anisotropy field
($H_{A}\approx$ 0.01 T) is much smaller than the exchange field ($H_{E}%
\approx$ 4.25 T) \cite{peng_magnetic_2020, calder_magnetic_2020}. CrPS\textsubscript{4} is therefore an excellent platform to study magnons
in controlled non-collinear spin textures because the moderate
spin-flop and spin-flip critical fields are accessible by standard lab
equipment.

Figure 1(b) shows the calculated band-edge (\(k=0\)) frequencies of the acoustic and optical
magnons of a bilayer of CrPS\textsubscript{4} with  easy axis along $z$ as a
function of ip magnetic fields using the parameters above. Fig. 1(c) sketches
the magnetization precession amplitudes for fields normal to the layers below the spin-flop
transition ($H<H_{\mathrm{spinflop}}$) in which the N\'{e}el vector remains
along $z\ $and the magnon modes carry opposite spins $+\hbar$/$-\hbar$, in the z-direction. Fig. 1(d) sketches the excitations of the canted spin texture at an ip magnetic
field below the spin-flip transition ($H<H_{\mathrm{E\perp}}$). The associated
magnons evolve from the zero-field spin up and down states with a net
magnetization along $y$ as indicated by the purple arrows.

\begin{figure}[!]
\centering
\includegraphics[width=\linewidth]{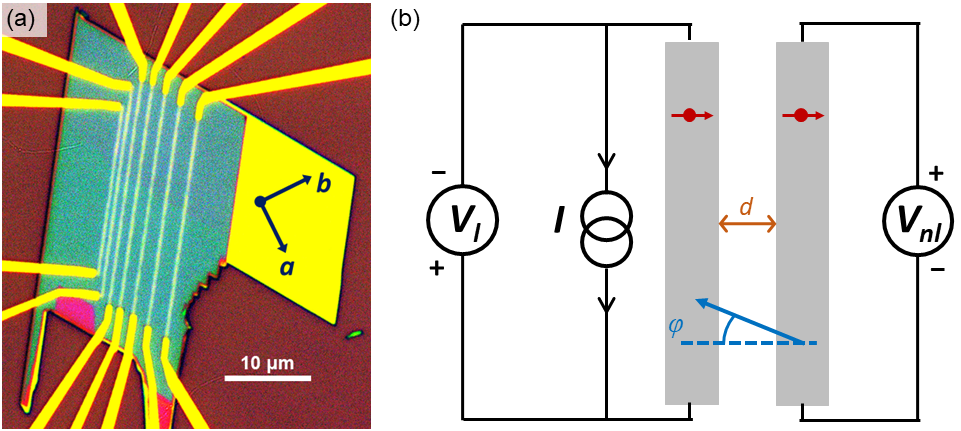}\caption{(a) Optical
micrograph of a transport device with 7 parallel Pt strips bonded by Ti/Au
leads on top of CrPS\textsubscript{4} film, where \textit{a} and \textit{b}
indicate the orientation.of the single crystal. (b) Electrical measurement
circuit, in which the red arrows indicate electrically active spins in the Pt
strips and $\varphi$ is an ip magnetic field angle.\ }%
\label{fig:Figure2}%
\end{figure}

We fabricated three devices by depositing multiple parallel Pt strips on
exfoliated CrPS\textsubscript{4} flakes with a thickness of $\sim$100 nm (see
figure \ref{fig:Figure2}(a)). We study both the local and non local
resistances as a function of magnitude and direction of an ip magnetic field.
We measure the magnetoresistance of a single Pt strip (SMR)
\cite{nakayama_spin_2013} as well as magnon transport and the spin Seebeck
effect non-locally by two Pt strips (see Fig. \ref{fig:Figure2}(b)).

Via the SMR we monitor the surface magnetization as a function of temperature,
ip external field, and bias current. The current $I$ in a Pt strip generates a
transverse spin current that when partially reflected at the Pt%
$\vert$%
CrPS\textsubscript{4} interface induces an additional current by the inverse
spin Hall effect, effectively reducing the electrical resistance. A
polarization of the spin-Hall spin current (red arrows in Fig.
\ref{fig:Figure2}(b)) parallel (normal) to the local moments of the magnet at
the interface, minimizes (maximizes) the dephasing by the exchange interaction
and therefore the electric resistance \cite{nakayama_spin_2013}.
An in-plane magnetic field $\mathbf{H}=H\mathbf{\hat{y}}$ (ip angle
$\varphi=0)$ cants the oop antiferromagnetic order by an angle
\begin{equation}
\theta_{\perp}=\arcsin\frac{H}{2H_{E}+H_{A}}%
\label{eq:thetaperp}
\end{equation}
with the $z$-axis. The electric resistance $R_{l}$ of a Pt wire along the
$x$-axis therefore should be maximal for $\theta_{\perp}=0$ and minimal for
$\theta_{\perp}=\pi/2,$ i.e. at and beyond the spin-flip transition. On the
other hand, magnon injection is most efficient when magnetic moments and
current-induced spins are parallel, maximizing the non-local resistance
$R_{nl}=V_{\text{\textrm{detector}}}/I_{\text{\textrm{injector}}}$.
$R_{nl}>0$ by defining the polarity of the voltage on the detector opposite to that
of the current in the injector (see figure \ref{fig:Figure2}(b)) .

The Joule heating by a charge current $I$ generates a temperature gradient
over the interface and in the magnet, generating a spin current and associated
inverse spin Hall voltage\ (spin Seebeck effect) in the injector as well as
the detector with associated local and non-local voltage signals.

We can separate the electrical and thermal signals by recording the first and
second harmonic responses to a current bias that oscillates with frequency
$\omega$. The first harmonic response reflects the Ohmic signal $V\sim I,$
while the thermal signals $V\sim I^{2}$\ appear at double frequency. Here we
focus on the linear response $R_{l/nl}^{\left(  1\omega\right)  }$, with a
brief discussion of $R_{l/nl}^{\left(  2\omega\right)  }$ in the Supplementary
Material (SM). The spin Hall effect and inverse spin Hall effect
dictate the following dependence on the ip field angle:
\begin{align}
R_{l}^{\left(  1\omega\right)  }  & =R_{l,0}^{\left(  1\omega\right)  }+\Delta
R_{l}^{\left(  1\omega\right)  }\sin^{2}\varphi\label{eq1}\\
R_{nl}^{\left(  1\omega\right)  }  & =\Delta R_{nl}^{\left(  1\omega\right)
}\cos^{2}\varphi
\end{align}
where $R_{0,l/nl}^{\left(  1\omega\right)  }$ are constant offsets and $\Delta
R_{l/nl}^{\left(  1\omega\right)  }$ are the strenghts of the signals that depend on the ip field 
angle $\varphi$ (defined in Fig. \ref{fig:Figure2}).  

We measured the local and non-local resistance in a liquid-He cryostat at
temperatures between 5 and 300 K as a function of an ip magnetic field up to
7.9 T and as a function of ip angle $\varphi$. Figure \ref{fig:Figure2} shows the
schematics of device D1 and the local and non-local measurement
configurations. The measurements on D1 were carried out on three different
pairs of Pt contacts with edge-to-edge distances of 330 nm, 420 nm and 780 nm.
On device D2, we measured the resistances for two different pairs ($\sim$300
nm and $\sim$450 nm) of Pt contacts. Results for device D3 with similar
flake thickness and contacts with an edge to edge spacing of 300 nm are
shown in the SM.

The time-dependent voltage responses may be expanded as $V(t)=R_{1}%
I(t)+R_{2}I^{2}(t)+\cdots$, where standard low frequency (7 Hz - 17 Hz)
lock-in techniques access the constants $R_{1}$ and $R_{2}$
\cite{cornelissen_long-distance_2015, bakker_interplay_2010}


\begin{figure}[ptb]
\centering
\includegraphics[width=0.93\linewidth]{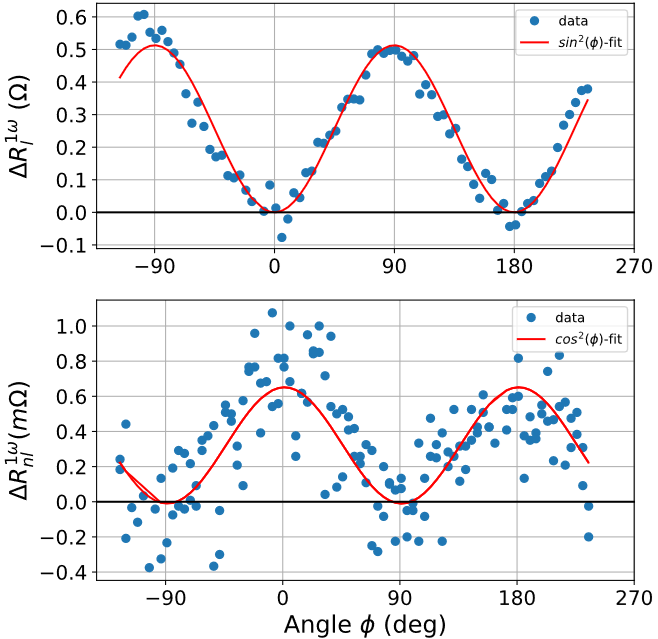}\caption{\textit{Top
panel: }Local resistance modulation $\Delta R_{l}^{\left(  1\omega\right)  }$
of the Pt strip as function of in-plane angle $\varphi$ (relative to the wire
normal) of an applied magnetic field of 7 T. The bias current is 60
$\mu$\textrm{A and} sample temperature is 24 K. The red curve is a fit by
$\sin^{2}\varphi$. A $5.97$ $\mathrm{k}\Omega$ offset resistance has been
subtracted. \textit{Bottom panel}: Simulataneously measured non-local
resistance $R_{nl}^{\left(  1\omega\right)  }$, fitted by $\cos^{2}\varphi$
(red curve).}
\label{fig:Figure3}%
\end{figure}

Fig. \ref{fig:Figure3} shows the observed $R_{l}^{\left(  1\omega\right)
}-R_{l,0}^{\left(  1\omega\right)  }\left(  \approx6\,\mathrm{k\Omega}\right)  $
of the injector contact as function of the direction of an ip magnetic field of 7 T
at AC current bias of 60 $\mathrm{\mu}$A and at $T=24$ K.
The observed $\varphi$ dependence agrees well with the model for the
SMR sketched above. $\Delta R_{l}^{\left(  1\omega\right)  }/R_{l,0}^{\left(
1\omega\right)  }\simeq10^{-4}$ is of the same order of magnitude as the SMR
of CrPS\textsubscript{4} in the oop configuration \cite{wu_magnetotransport_2022}
and that of other magnetic materials, which is a strong
indication of an efficient interface exchange coupling and a large spin-mixing conductance.
\begin{figure}[ptb]
\centering
\includegraphics[width=0.93\linewidth]{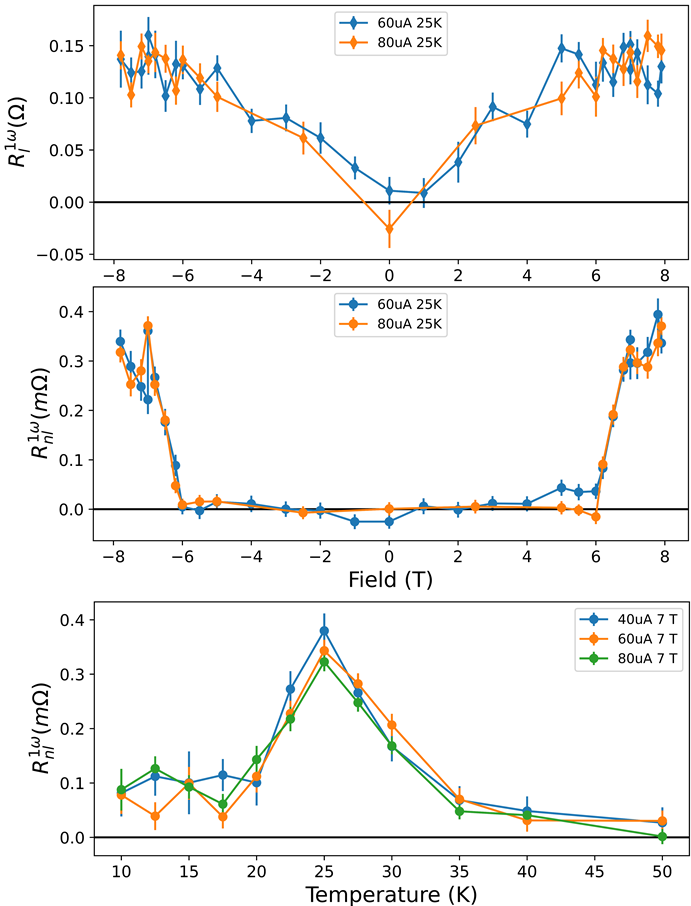}\caption{
Field- and temperature-dependent results on device D2, \textit{Top}: Field
dependence of $R_{l}^{1\omega}$ at different bias currents at 25 Kelvin.
\textit{middle}: Same for $R_{nl}^{1\omega}$. \textit{Bottom}: Temperature
dependence of $R_{nl}^{1\omega}$ at 7 T.}%
\label{fig:Figure4}%
\end{figure}

The modulation $\Delta R_{l}^{\left(  1\omega\right)  }$ at $T=20$ K in Fig.
\ref{fig:Figure4}(a) as a function of magnetic field strength agrees also with
expectations, while the lack of a bias-current dependence confirms that we are
in the linear response regime. We observe saturation at fields $>$ 6T (see Fig. \ref{fig:Figure4}), which corresponds to the onset of the spin-flip state.


At fields of several tesla, the SMR decreases with temperature
but persists above $T_{N}$ and even up to room temperature (not shown), which
is consistent with reports for CrPS\textsubscript{4} \cite{wu_magnetotransport_2022} and the van der Waals material Cr$_{2}$Ge$_{2}%
$Te$_{6}$ \cite{zhu_interface-enhanced_2022}. The robust SMR can possibly be ascribed
to a $T_{N}$ that is enhanced by the interface spin orbit coupling and/or a
paramagnetic SMR by a field-induced magnetization \cite{oyanagi_paramagnetic_2021}.

We now focus on the non-local signal $R_{nl}^{\left(  1\omega\right)  }$
plotted in the lower panel Fig. \ref{fig:Figure3} for current bias of 60 $\mu
$\textrm{A}, $H=7$ T, $T=24$ K (measured together with $R_{l}^{\left(1\omega\right) }$).
$\Delta R_{l}^{\left(1\omega\right) }\simeq 0.6\,\mathrm{m\Omega}$ is about 30 times smaller than
that of Pt|YIG (thickness of 200 nm) \cite{cornelissen_long-distance_2015}. $R_{nl}^{\left(  1\omega\right)  }$ is maximum
(minimum) at $\phi=0^{\circ}$ ($\phi=90^{\circ}$) which reflects the angular dependence of the spin injection and detection efficiencies by the spin Hall effects in Pt.

Fig. \ref{fig:Figure4}\ reveals a remarkable dependence of the magnon
transport on magnetic field. At $H$ $\leq$ 6 T, no $R_{nl}^{\left(
1\omega\right)  }$ is observed within the experimental uncertainty. At fields $>$
6 T, $R_{nl}^{\left(  1\omega\right)  }$ increases sharply and appears to
saturate at fields $>$ 7 T. This rapid increase correlates with the saturation of the bulk
magnetization (see SM) and is therefore associated with the
spin-flip transition from a canted antiferromagnetic (AFM) to a collinear ferromagnetic (FM) state (see Fig. 1(c\&d)).

The non-local resistance at current bias of 60 $\mathrm{\mu}$\textrm{A} and at
7 T in Fig. \ref{fig:Figure4} is non-monotonous, with a maximum around 25 K. We attribute this to two effects.
On one hand, the critical fields for the spin-flip transition decreases with temperature. At low temperatures and 7 T, the sample is still in the canted-AFM phase. The sharp increase in $R_{nl}%
^{\left(  1\omega\right)  }$ coincides with the formation of the saturated FM phase at $T\sim25$ K. Moreover, the equilibrium magnon density and resulting magnon conductivity increase with temperature \cite{cornelissen_temperature_2016, cornelissen_magnon_2016}. 
At larger temperatures, the magnetization and $R_{nl}^{\left(  1\omega\right)  }$ decrease and vanish at $T_{N}$.

\begin{figure}[ptb]
\centering
\includegraphics[width=0.95\linewidth]{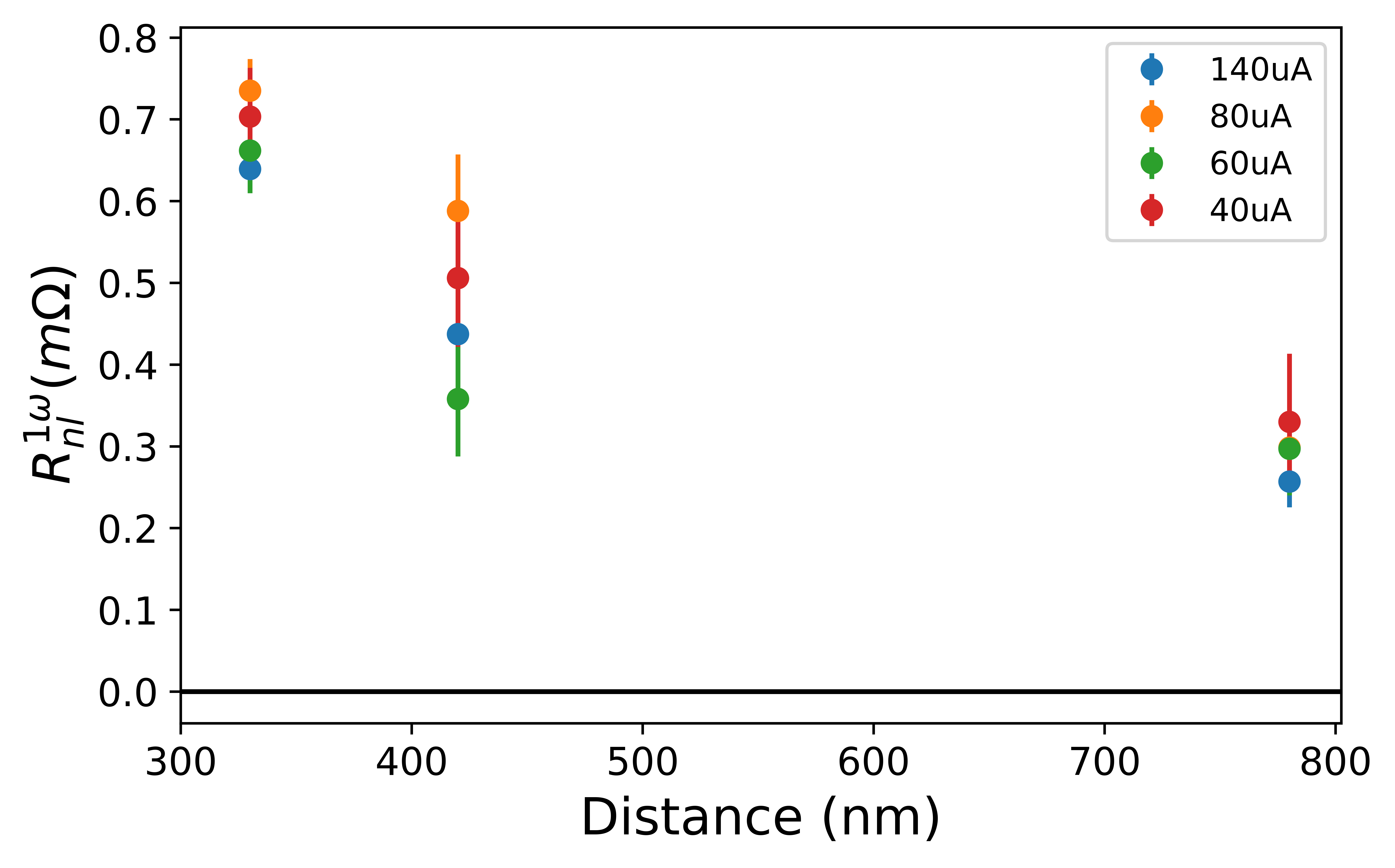}\caption{Non-local
resistance as function of distance $d$ between the injector and detector
contacts. For different bias currents at sample temperature of 24 K, $R_{nl}^{\left(  1\omega\right)  }\sim1/d$ is shown for three different Pt strip pairs on device D1 with $d$ being equal to 330 nm, 420 nm and 780 nm, respectively.  }%
\label{fig:Figure5}%
\end{figure}

Further, we assess the magnon transport in CrPS\textsubscript{4} by measuring as a
function of distance $d$ between the Pt contacts as shown in Fig.
\ref{fig:Figure5}. The absence of a systematic dependence on the current bias
again confirms that we operate in the linear response regime. The model of diffusive magnon transport in YIG leads to as decay of $R_{nl}^{\left(  1\omega\right)  }$ with increasing $d$ as a function of the magnon diffusion length \(\lambda\). For efficient spin injection at the Pt$|$YIG interface, this described by \cite{cornelissen_long-distance_2015}:
\begin{equation}%
R_{nl}^{\left(  1\omega\right)  }=\frac{C}{\lambda}\frac{\exp{\left( d/\lambda\right)}}{1-\exp{\left( 2d/\lambda\right)}},
\label{eq:Ludo}
\end{equation}
where $C$ is a constant. Since we observe an algebraic
$R_{nl}^{\left(  1\omega\right)  }\sim1/d$ rather than exponential dependence, magnon transport over the length scales $d\leq$ 1$\,\mathrm{\mu
m}$ is Ohmic \cite{cornelissen_long-distance_2015}, i.e. purely diffusive while
magnon decay sets in at larger distances only. 

The abrupt field dependence of the non-local resistance differs sharply from
the linear dependence of the SMR (top panes of Fig. 4) that indicates a surface magnetization proportional to
a static magnetic susceptibility. A surprise of the present study is the absence 
of non-local transport in the non-collinear phase. This behavior
is markedly different from previous studies of transport that were carried out
with magnetic fields parallel to the N\'{e}el vector, including the spin-flop
transition \cite{lebrun_anisotropies_2019,hahn_comparative_2013,das_anisotropic_2022}. However, the associated theories do not address the present configuration either.

The magnon band edges of CrPS\textsubscript{4} in the canted phase as plotted
in Fig. 1(b) diagonalize the classical spin Hamiltonian with eigenfrequencies
\cite{gurevich_magnetization_1996} at low fields \(\ H\leq H_{\mathrm{E\perp}}=2H_{E}+H_{A}\)

\begin{equation}%
\begin{split}
&\omega_{+}=\gamma\sqrt{(2H_{E}\sin^{2}\theta_{\perp}+H_{A}\cos^{2}%
\theta_{\perp})(2H_{E}+H_{A})}\\
&\omega_{-}=\gamma\sqrt{H_{A}(2H_{E}+H_{A})\cos^{2}\theta_{\perp}}%
%
\end{split}
\end{equation}
and at high fields \(H>H_{\mathrm{E\perp}}=2H_{E}+H_{A}\)%
\begin{equation}%
\begin{split}
&\omega_{+}=\gamma\sqrt{(H-H_{A})H}\\
&\omega_{-}=\gamma\sqrt{(H-2H_{E})(H-2H_{E}-H_{A})},%
\end{split}
\end{equation}where $\theta_{\perp}$ is defined in equation \ref{eq:thetaperp} and includes the external ip field $H$. 
The small anisotropy causes the low frequencies of the acoustic modes
that at high magnetic fields and low temperatures are dominantly populated. 

The collinear ferromagnetic phase above $H_{\mathrm{E\perp}}$
can be treated by a two-mode linearized Boltzmann equation similar to YIG,
while the large SMR\ implies that interfaces are transparent, so we may expect at high non-local signal at \(H>H_{E\perp}\). As the Pt contacts do not inject or detect spin polarizations in the z-direction.
At zero canting angle ($\theta_{\mathrm{\perp}}$), the spin current injected by the Pt contacts is fully absorbed by the antiferromagnet in the form of a spin transfer torque to the magnetic sublattices, while magnon injection and \(R_{nl}^{\left(  1\omega\right)  }\) vanish.
With increasing canting angle, the magnon injection efficiency increases proportional with the induced net magnetization. However, the  exchange interaction in a non-collinear configuration also increasingly affects the non-local signal by reducing the magnon decay length (P. Tang, in preparation). 
In the collinear phase both magnon injection and magnon transport do not prevent the non-local signal. The abruptness of the observed onset of non-local transport at the spin-flip transition field is surprising, however. The abrupt increase of non-local resistance near the spin flip transition may be caused by the combined effects of enhanced magnon injection into the low energy magnon branch and the sudden suppression of magnon relaxation when the system approaches the FM state.

Summarizing, we report non-local spin transport in a van der Waals magnet, to
the best of our knowledge for the first time. The spin conduit is the
electrically insulating antiferromagnet CrPS\textsubscript{4} with
perpendicular anisotropy. We focus on a configuration that has escaped attention even in conventional antiferromagnets, with an in-plane magnetic field
normal to the Pt spin injector and detector that tilts the antiparallel spins into
the plane. Surprisingly, we do not observe spin transport in the
non-collinear phase. At the critical field that forces
the transition to a collinear ferromagnetic phase, we observe an abrupt
increase of the non-local spin signal over distances that exceed a micron.
These results herald the potential of 2D van der Waals magnets for scalable
magnonic circuits. 

\begin{acknowledgments}
We acknowledge the technical support from J.\ G.\ Holstein, H.\ Adema T.\ Schouten,  H.\ H. de Vries and F.\ H. van der Velde.
We acknowledge the financial support of the Zernike Institute for Advanced Materials and the European Union’s Horizon 2020 research and innovation program under Grant Agreement No. 785219 and No. 881603 (Graphene Flagship Core 2 and Core 3).
This project is also financed by the NWO Spinoza prize awarded to BJW by the NWO and
has received funding from the European Research Council (ERC) under the European Union’s 2DMAGSPIN
(Grant agreement No. 101053054). GB acknowledges funding by JSPS Kakenhi grant no. 19H00645.	
\end{acknowledgments}

\bibliography{Reference}

\end{document}